\documentclass[longauth,oldversion]{aa}

\usepackage{txfonts}
\usepackage[final]{graphicx}
\graphicspath{{./Input/}}

\usepackage{amstext,amssymb}
\usepackage{natbib}
\bibpunct{(}{)}{;}{a}{}{,}

\usepackage{color}

\newcommand{\Rstar}{\ensuremath{R_\star}}
\newcommand{\Macc}{\ensuremath{\dot{M}_\text{acc}}}
\newcommand{\Rin}{\ensuremath{R_\text{in}}}
\newcommand{\Rout}{\ensuremath{R_\text{out}}}
\newcommand{\Vline}{\ensuremath{V_\text{line}}}
\newcommand{\Vcont}{\ensuremath{V_\text{cont}}}
\newcommand{\Fline}{\ensuremath{F_\text{line}}}
\newcommand{\Fcont}{\ensuremath{F_\text{cont}}}
\newcommand{\Av}{\ensuremath{A_V}}
 
\newcommand{\microns}{\ensuremath{\mu\text{m}}}
\newcommand{\nm}{\ensuremath{\text{nm}}}
\newcommand{\mas}{\ensuremath{\text{mas}}}
\newcommand{\AU}{\ensuremath{\text{AU}}}

\newcommand{\degrees}{\ensuremath{^{\circ}}}
\newcommand{\yr}{\ensuremath{\text{yr}}}
\newcommand{\s}{\ensuremath{\text{s}}}
\newcommand{\ms}{\ensuremath{\text{ms}}}
\newcommand{\K}{\ensuremath{\text{K}}}
\newcommand{\pc}{\ensuremath{\text{pc}}}
\newcommand{\Mag}{\ensuremath{\text{mag}}}
\newcommand{\kms}{\ensuremath{\text{km\,s}^{-1}}}
\newcommand{\Rsun}{\ensuremath{R_\odot}}

\newcommand{\Msun}{\ensuremath{M_\odot}}
\newcommand{\MsunPyr}{\ensuremath{\Msun\,\yr^{-1}}}
\newcommand{\Ha}{\ensuremath{\text{H}\alpha}}
\newcommand{\Hb}{\ensuremath{\text{H}\beta}}
\newcommand{\Bra}{\ensuremath{\text{Br}\alpha}}
\newcommand{\Brg}{\ensuremath{\text{Br}\gamma}}
\newcommand{\lBrg}{\ensuremath{2.1656\,\microns}}
\newcommand{\mwc}{\object{MWC\,297}}
\newcommand{\calib}{\object{HD\,177756}}

\begin{document}
\title{Disk and wind interaction in the young stellar object
  \mwc\ spatially resolved with VLTI/AMBER\thanks{Based on
    observations collected at the European Southern Observatory during
    commissioning, Chile}}
\titlerunning{Disk and wind interaction in \mwc\ resolved
  with VLTI/AMBER}
\author{F.~Malbet\inst{1}
  \and M.~Benisty\inst{1}
  \and W.J.~de Wit\inst{1}
  \and S.~Kraus\inst{2}
  \and A.~Meilland\inst{3}
  \and F.~Millour\inst{1,4}
  \and E.~Tatulli\inst{1}
  \and J.-P.~Berger\inst{1}
  \and O.~Chesneau\inst{3}
  \and K.-H.~Hofmann\inst{2}
  \and A.~Isella\inst{5,12}
  \and A.~Natta\inst{5}
  \and R.~Petrov\inst{4}
  \and T.~Preibisch\inst{2}
  \and P.~Stee\inst{3}
  \and L.~Testi\inst{5}
  \and G.~Weigelt\inst{2}
  \and P.~Antonelli\inst{3}
  \and U.~Beckmann\inst{2}
  \and Y.~Bresson\inst{3}
  \and A.~Chelli\inst{1}
  \and G.~Duvert\inst{1}
  \and L.~Gl\"uck\inst{1}
  \and P.~Kern\inst{1}
  \and S.~Lagarde\inst{3}
  \and E.~Le~Coarer\inst{1}
  \and F.~Lisi\inst{5}
  \and K.~Perraut\inst{1}
  \and S.~Robbe-Dubois\inst{4}
  \and A.~Roussel\inst{3}
  \and G.~Zins\inst{1}
  \and M.~Accardo\inst{5}
  \and B.~Acke\inst{1,13}
  \and K.~Agabi\inst{4}
  \and B.~Arezki\inst{1}
  \and E.~Aristidi\inst{4}
  \and C.~Baffa\inst{5}
  \and J.~Behrend\inst{2}
  \and T.~Bl\"ocker\inst{2}
  \and S.~Bonhomme\inst{3}
  \and S.~Busoni\inst{5}
  \and F.~Cassaing\inst{6}
  \and J.-M.~Clausse\inst{3}
  \and J.~Colin\inst{3}
  \and C.~Connot\inst{2}
  \and A.~Delboulb\'e\inst{1}
  \and T.~Driebe\inst{2}
  \and M.~Dugu\'e\inst{3}
  \and P.~Feautrier\inst{1}
  \and D.~Ferruzzi\inst{5}
  \and T.~Forveille\inst{1}
  \and E.~Fossat\inst{4}
  \and R.~Foy\inst{7}
  \and D.~Fraix-Burnet\inst{1}
  \and A.~Gallardo\inst{1}
  \and S.~Gennari\inst{5}
  \and A.~Glentzlin\inst{3}
  \and E.~Giani\inst{5}
  \and C.~Gil\inst{1}
  \and M.~Heiden\inst{2}
  \and M.~Heininger\inst{2}
  \and D.~Kamm\inst{3}
  \and D.~Le Contel\inst{3}
  \and J.-M.~Le Contel\inst{3}
  \and B.~Lopez\inst{3}
  \and Y.~Magnard\inst{1}
  \and A.~Marconi\inst{5}
  \and G.~Mars\inst{3}
  \and G.~Martinot-Lagarde\inst{8,14}
  \and P.~Mathias\inst{3}
  \and J.-L.~Monin\inst{1}
  \and D.~Mouillet\inst{1,15}
  \and D.~Mourard\inst{3}
  \and P.~M\`ege\inst{1}
  \and E.~Nussbaum\inst{2}
  \and K.~Ohnaka\inst{2}
  \and J.~Pacheco\inst{3}
  \and F.~Pacini\inst{5}
  \and C.~Perrier\inst{1}
  \and P.~Puget\inst{1}
  \and Y.~Rabbia\inst{3}
  \and S.~Rebattu\inst{3}
  \and F.~Reynaud\inst{9}
  \and A.~Richichi\inst{10}
  \and M.~Sacchettini\inst{1}
  \and P.~Salinari\inst{5}
  \and D.~Schertl\inst{2}
  \and W.~Solscheid\inst{2}
  \and P.~Stefanini\inst{5}
  \and M.~Tallon\inst{7}
  \and I.~Tallon-Bosc\inst{7}
  \and D.~Tasso\inst{3}
  \and J.-C.~Valtier\inst{3}
  \and M.~Vannier\inst{4,11}
  \and N.~Ventura\inst{1}
  \and M.~Kiekebusch\inst{11}
  \and F.~Rantakyr\"o\inst{11}
  \and M.~Sch\"oller\inst{11}
}             
 
\offprints{F.~Malbet\\ email: \texttt{<Fabien.Malbet@obs.ujf-grenoble.fr>}}
\institute{
  Laboratoire d'Astrophysique de Grenoble, UMR 5571 Universit\'e Joseph
  Fourier/CNRS, BP 53, F-38041 Grenoble Cedex 9, France
  \and Max-Planck-Institut f\"ur Radioastronomie, Auf dem H\"ugel 69,
  D-53121 Bonn, Germany
  \and Laboratoire Gemini, UMR 6203 Observatoire de la C\^ote
  d'Azur/CNRS, Avenue Copernic, 06130 Grasse, France
  \and Laboratoire Universitaire d'Astrophysique de Nice, UMR 6525
  Universit\'e de Nice/CNRS, Parc Valrose, F-06108 Nice cedex 2, France
  \and INAF-Osservatorio Astrofisico di Arcetri, Istituto Nazionale di
  Astrofisica, Largo E.~Fermi 5, I-50125 Firenze, Italy
  \and ONERA/DOTA, 29 av de la Division Leclerc, BP 72, F-92322
  Chatillon Cedex, France 
  \and Centre de Recherche Astronomique de Lyon, UMR 5574 Universit\'e
  Claude Bernard/CNRS, 9 avenue Charles Andr\'e, F-69561 Saint Genis
  Laval cedex, France
  \and Division Technique INSU/CNRS UPS 855, 1 place Aristide
  Briand, F-92195 Meudon cedex, France
  \and IRCOM, UMR 6615 Universit\'e de Limoges/CNRS, 123 avenue Albert
  Thomas, F-87060 Limoges cedex, France
  \and European Southern Observatory, Karl Schwarzschild Strasse 2,
  D-85748 Garching, Germany
  \and European Southern Observatory, Casilla 19001, Santiago 19,
  Chile
  \and Dipartimento di Fisica, Universit\`a degli Studi di Milano, Via
  Celoria 16, I-20133 Milano, Italy
  \and Instituut voor Sterrenkunde, KULeuven, Celestijnenlaan 200B,
  B-3001 Leuven, Belgium 
  \and \emph{Present affiliation:} Observatoire de la Côte d'Azur -
  Calern, 2130 Route de l'Observatoire , F-06460 Caussols, France
  \and \emph{Present affiliation:} Laboratoire Astrophysique de
  Toulouse, UMR 5572 Universit\'e Paul Sabatier/CNRS, BP 826, F-65008
  Tarbes cedex, France 
}
\date{Received date; accepted date}
%


\abstract{The young stellar object \mwc\ is an embedded B1.5Ve star
  exhibiting strong hydrogen emission lines and a strong near-infrared
  continuum excess. This object has been observed with the VLT
  interferometer equipped with the AMBER instrument during its first
  commissioning run.  VLTI/AMBER is currently the only near infrared
  interferometer which can observe spectrally dispersed visibilities.
  \mwc\ has been spatially resolved in the continuum with a visibility
  of $0.50^{+0.08}_{-0.10}$ as well as in the \Brg\ emission line
  where the visibility decrease to a lower value of $0.33\pm0.06$.
  This change in the visibility with the wavelength can be interpreted
  by the presence of an optically thick disk responsible for the
  visibility in the continuum and of a stellar wind traced by the
  \Brg\ emission line and whose apparent size is 40\% larger.  We
  validate this interpretation by building a model of the stellar
  environment that combines a geometrically thin, optically thick
  accretion disk model consisting of gas and dust, and a
  latitude-dependent stellar wind outflowing above the disk surface.
  The continuum emission and visibilities obtained from this model are
  fully consistent with the interferometric AMBER data. They agree
  also with existing optical, near-infrared spectra and other
  broad-band near-infrared interferometric visibilities. We also
  reproduce the shape of the visibilities in the \Brg\ line as well as
  the profile of this line obtained at an higher spectral resolution
  with the VLT/ISAAC spectrograph, and those of the \Ha\ and \Hb\
  lines. The disk and wind models yield a consistent inclination of
  the system of approximately 20\degrees. A picture emerges in which
  \mwc\ is surrounded by an equatorial flat disk that is possibly
  still accreting and an outflowing wind which has a much higher
  velocity in the polar region than at the equator.  The VLTI/AMBER
  unique capability to measure spectral visibilities therefore allows
  us for the first time to compare the apparent geometry of a wind
  with the disk structure in a young stellar system.
}
\maketitle
\section{Introduction}
\label{sect:intro}

Pre-main sequence stars in the intermediate mass range, called Herbig
Ae and Be stars (HAeBe), 
are observed to be surrounded by circumstellar material which reveals
itself by discrete emission lines and by continuous excess emission in
the spectral energy distribution (SED). The spatial distribution of
this material however has been subject to debate: both
geometrically flat disk models and spherically symmetric envelope
models can reproduce the observed SED. Using viscous accretion disk
models, \citet{Hil1992} proposed a disk-like geometry for the
circumstellar material that however needs a central cavity to fit the
near-infrared (NIR) excess. Considerations on the physical reality of these
central cavities \citep{Ken1993} and successful fits to HAeBe SEDs by
dust envelopes \citep{Ber1992,Fran1994} or composite envelope-disk
models \citep{Mir1999} introduced the controversy on the geometry of
the circumstellar material.

A break-through occurred with high resolution interferometric
observations of HAeBe stars first presented by \citet{MST2001} and
\citet{Tut2001}. These observations revealed that the geometries
observed in the near-infrared bands were closer to either ring-like or
spherically symmetric rather than disk-like.  New models introducing
passive, star-irradiated circumstellar disks with puffed up inner rims
were found to be consistent with both the interferometric and
photometric observations \citep{Nat2001,DDN2001} although no
simultaneous fits were performed. In this scenario the inner part of
the circumstellar disk edge puffs up due to direct irradiation by the
central star, roughly equivalent to the ring-like structures found
with interferometry. The location of this inner rim is near the dust
sublimation radius, and the subsequent (dust-free) inner holes are
much larger than the ad-hoc inner holes proposed in \citet{Hil1992}.
Recent studies show that there seems to exist a correlation between
the distance of the inner rim and the luminosity of the central star
\citep{MM2002}.  This relation holds well for the Herbig Ae and
late type Be stars, but breaks down for the most luminous HBe
stars \citep{Eis2004,Mon2005}. The latter authors in fact demonstrate
that the early-type HBe have inner rims that are too close to the star
with respect to the star's luminosity.  The HBe seem to better match
the ``classical'' viscous accretion disk model similar to the one
initially applied by \citet{Hil1992}. A marked difference however is
that now the inner hole should be filled with optically thick gas to
effectively shield the dust from destruction at relatively small
distances from the inner rim, assuming that dust is what is observed in
interferometry. Additional strong evidence for a different disk
character near the early-type HBe stars is found in the
spectro-polarimetric fingerprint of \Ha\ lines \citep{Vin2002}. These
authors prefer to draw the analogy with the geometrically flat gaseous
disks present near the classical Be stars.

The geometry of circumstellar material near HAeBe stars thus seems to
differ between the early-type and late-type members of the group,
which is not surprising given the increasing interaction between star
and disk for the early type stars.  For the HAe stars a successful
working model exists, while on the other hand, a disk structure near
the HBe stars and their intricate star-disk interactions still escape
a good understanding. In this study we present high spatial
resolution, intermediate spectral resolution interferometric
observations of the early-type Herbig Be star \mwc using VLTI/AMBER.
This enigmatic star appeared in the original HAeBe list compiled by
\citet{Her1960}. The star displays a strong emission line spectrum and
the character of the underlying photosphere was revealed in the
detailed study by \citet{Drew1997} to be B\,1.5Ve. Its rather well
determined stellar parameters and its high NIR luminosity render this
star the perfect target to investigate in detail the geometry of the
circumstellar material near the early type HBe stars.

The paper is organized as follows. Section~\ref{sect:obs} presents the
new observations made with AMBER and ISAAC. In Sect.~\ref{sect:amber},
we describe the AMBER data processing to produce a reliable result
presented in Sect.~\ref{sect:results} for the visibilities both in the
continuum and in the \Brg\ line. In Sect.~\ref{sect:models} we present
an attempt to model the environment of \mwc\ with a disk and a wind.
This model and the consequences are discussed in
Sect.~\ref{sect:discussion} and summarized in Sect.~\ref{sect:summ}.

\section{Observations}
\label{sect:obs}
\subsection{AMBER observations}

\begin{table}[t]
  \centering
  \caption{AMBER observation log from 31 May 2004 (UT).}
  \label{tab:log}
  \smallskip
  \begin{tabular}{lclllll}
  \hline
  Star      & DIT   &UT     &UT     &Nb     &u    &v   \\
            & ($\ms$) &start&end    &exp.   &(m)  &(m) \\
  \hline
  \mwc\     & 107 &06:02    &06:13  &18     &28.19 &33.57 \\
  \mwc\     & 31  &06:26    &06:37  &40     &29.93 &33.37 \\
  \calib\ & 31  &08:10    &08:18  &30     &32.92 &32.85 \\
  \calib\ & 107 &08:21    &08:33  &14     &33.20 &32.69 \\
  \hline
  \end{tabular}
\end{table}

\mwc\ was observed on 31 May 2004 during the second night of the first
commissioning run of the AMBER instrument on the UT2-UT3 (47\,m)
baseline of the \emph{Very Large Telescope Interferometer} (VLTI).
AMBER is the VLTI beam combiner operating in the near-infrared
\citep{Pet2003}.  The instrument is based on spatial filtering with
fibers and multiaxial fringe coding (i.e.\ AMBER is combining the
beams at an angle which results in fringes modulated in the spatial
direction). The interferometric beam is anamorphized perpendicularly
to the fringe coding in order to be injected into the slit of a
spectrograph. The instrument can operate at spectral resolutions up to
10,000 and efficiently deliver spectrally dispersed visibilities.

\mwc\ was measured in the [1980,2230nm] spectral range in the
\texttt{MR-K} spectral mode (spectral resolution of 1500) with 2
elementary detector integration times (DIT) of $31\,\ms$ and
$107\,\ms$. Table \ref{tab:log} gives the log of the AMBER
observations. \calib, whose spectral type is B\,9V, was used to
calibrate the visibilities. Its diameter is $0.60\pm0.06$ as computed
by the ASPRO \texttt{searchCalib}\footnote{ASPRO is available at
  \texttt{http://mariotti.fr}} tool developed at the \emph{Jean-Marie
  Mariotti Center}.

The observations of \mwc\ were carried out under specific conditions
since it was the first AMBER commissioning run on the VLTI. The bright
\Brg\ line of \mwc\ has been originally observed in order to perform a
spectral calibration of AMBER. Detailed analysis of the commissioning
data from this run and later ones has shown that the optical train of
the UT telescopes were affected by non-stationary high-amplitude
vibrations.  Because of the small number of observations on \mwc, and
because the amplitude of the vibrations might undergo rapid
variations, the calibration of our measurements must therefore be
regarded with care.  In order to investigate the errors on the
visibility measurements and check their consistency, we have used
different data reduction methods and different data selection schemes
which are described below. However, we would like to point out that
the vibrations do not impact the spectral dependence of the visibility
since they affect the spectral range as a whole.

\subsection{ISAAC observations}

\mwc\ was observed in service mode on 13 July 2004 with the ESO VLT UT1
telescope under modest seeing conditions ($\sim 1.5\arcsec$ in the
visual). The ISAAC near-infrared spectrograph was employed in the
short wavelength medium resolution mode with a $0.3\arcsec$ wide slit.
This instrument setup delivered a resolution of
$\lambda/\Delta\lambda\sim8900$ at the \Brg\ wavelength. The raw data
were flat-fielded, wavelength-calibrated and corrected for telluric
absorption using standard techniques and observations taken from the
ISAAC calibration plan by the ESO staff astronomers. A detailed
account of the observations and the data reduction is given in
\citet{GLeaPrep}.

\subsection{Existing photometric and interferometric data sets}

Extensive photometric and interferometric data exist for \mwc.
Broad-band photometric data were collected and presented by
\citet{Pez1997}. This data set consists of UBVRI from \citet{Ber1988},
JHKLMN from \citet{Ber1992}, and Q-band data from \citet{Sim1974}. In
the mm/submm wavelength regime (0.35-1.3\,mm) the dust continuum
measurements are taken from \citet{Man1994}.  At radio wavelengths
(6\,cm), \mwc\ has been observed by \citet{Drew1997}.  Existing NIR
interferometric data for \mwc\ consist of two sets. IOTA H-band
continuum data were presented by \citet{MST2001}, and PTI K-band
continuum data were published by \citet[][ upper limits only]{Eis2004}.

\section{AMBER data processing}
\label{sect:amber}


\subsection{Raw data}

\begin{figure}[t]
  \centering
  \includegraphics[width=0.9\hsize]{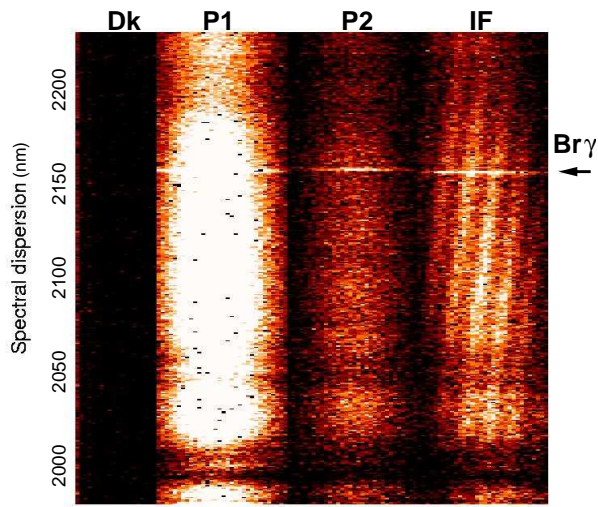}
  \caption{Short-exposure image of the \mwc\ signal on the AMBER
    detector. X-axis corresponds to the spatial extension of the beams
    and Y-axis to the wavelength.  First column (Dk) corresponds to
    the dark, the second (P1) and third (P2) ones are the beams from
    the first and second telescope resp., and finally the last column
    (IF) shows the fringes obtained by superposition of the two beams.
    The bright row is the \Brg\ line at $2165.6\,\nm$.}
\label{fig:rawdata}
\end{figure}
AMBER follows the standard data flow system implemented at ESO/VLT.
During data acquisition, the software records the images of the
spectrally dispersed fringes as well as those of the telescope beams.
Figure \ref{fig:rawdata} shows a typical image retrieved from the
detector subsystem\footnote{As a matter of fact, this image is not
  exactly how it appears on the detector, since 
  the four columns come from more widely separated regions of the
  detector.}.  From the left to the right, the first column contains
an estimation of the dark, the second and third ones are the
spectrophotometric measurements from each telescope in order to
evaluate their respective contribution to the interferogram, and the
last one is the spectrally dispersed fringe pattern, corresponding to
the interference between the two telescope beams.  Each exposure (or
file) consists of one hundred of these elementary frames.

One can notice the bright line near the top of the figure and the dark
lines at the bottom. They correspond  to the \Brg\
emission line at \lBrg\ and to the telluric absorption lines
around 2.05 and 2.08\,\microns\ respectively.

\subsection{Data reduction}
\label{sect:reduction}

Since we are still in an early stage of AMBER observations, we decided
to use two different methods to extract the raw visibilites. We
present and discuss them briefly.
\begin{itemize}
\item The standard data reduction method developed and optimized for
  AMBER is called P2VM for Pixel-To-Visibilities Matrix
  \citep{Mil2004,TatPrep}. The P2VM is a linear matrix method which
  derives raw visibilities from AMBER data for each spectral channel.
  The P2VM is computed after an internal calibration procedure which
  is performed every time the instrument configuration changes.  The
  complex coherent fluxes\footnote{Coherent flux is the degree of
    complex coherence not yet normalized by the flux.} are given by
  the product of the fluxes measured on each pixel of the detector by
  this P2VM.  We estimate the visibility for each spectral channel by
  integrating the square amplitude of the complex coherent fluxes over
  a certain number of frames and then by normalizing it by the total
  flux obtained for the same set of frames. The number of frames
  used can vary between 1 and the total number of frames in one
  exposure. This parameter is called hereafter the binning. 
\item A Fourier transform (FT) technique \citep[see][ appendix
  A]{Ohn2003} has been used. The advantage is that it does not depend
  on the internal calibration of the instrument, described above.  To
  enhance the signal, we applied a sliding average of seven spectral
  channels which has no effect on the continuum visibilities, but
  allows us to retrieve only upper limits of the visibility in the
  \Brg\ line. The visibility is computed using power spectral
  densities averaged over a variable number of frames.
\end{itemize}
Both methods compute squared visibilities and therefore an important
step is to subtract the visibility bias, i.e.\ the part of the
visibility due to photon and read-out noises. Because of the presence
of vibrations, we are not completely sure that this bias is perfectly
estimated and therefore we use very conservative errors.

\subsection{Data selection}
\label{sect:selection}

Because of the specific conditions of our observations, one critical
step in the processing is the selection of the best data within the
whole set. We based our selection of individual frames or exposures on
the value of the visibility signal-to-noise ratio (SNR).

We used two different methods of selection based either on exposure
or frame selection.
\begin{itemize}
\item The selection over the \emph{exposures} consists of selecting a
  certain fraction of the best exposures. The visibilities for each
  exposure are computed using all frames. The final value is the
  average of the visibilities over the selected exposures weighted by
  the statistical errors.
\item The selection over the \emph{frames} consists of selecting a
  certain fraction of the best frames within each exposures acquired on
  the object. The final value of the visibility is the average of
  the visibilities of all selected frames weighted by their errors.
\end{itemize}
The exposure selection has been used only with the P2VM method whereas
the frame selection has been used with both reduction methods. In all
cases, the selection has been applied to the object and its calibrator
and the resulting calibrated visibilities are summarized in Table
\ref{tab:V}.
\begin{table*}[t]
 {
  \begin{center}
    \caption[]{Calibrated visibilities obtained with different algorithms and
      data selection.}
    \begin{tabular}{lllcccc}
      \hline
      \hline
      Algorithm &Selection   &DIT (\ms) &\Vline &\Vcont    \\
      \hline
      P2VM &35\% best exposures &107 &$0.31\pm0.02$ &$0.58\pm0.03$  \\
      P2VM &25\% best exposures &31  &$0.33\pm0.03$ &$0.57\pm0.05$  \\
      P2VM &10\% best frames    &107 &$0.27\pm0.02$ &$0.41\pm0.04$  \\
      P2VM &10\% best frames    &31  &$0.36\pm0.03$ &$0.57\pm0.06$  \\
      FT   &10\% best frames    &107 &$\leq$0.31    &$0.39\pm0.05$  \\
      FT   &10\% best frames    &31  &$\leq$0.39    &$0.45\pm0.05$  \\
      \hline
 \end{tabular}
\label{tab:V}
\end{center}
}
\end{table*}

The threshold for the selection process is an important parameter.
Selecting with a strong criteria (e.g. $\leq 5\%$) would not provide
reliable statistics for the biases on the square visibility.
Selection with a soft criteria (e.g above $\geq 50\%$) may retain
influences by a reduced fringe contrast due to telescope vibrations.
Therefore we chose 10\% for the frame selection, and 25\% (resp.\
35\%) for the exposure selection of the $31\,\ms$ (resp.\ $107\,\ms$)
data set because of the small number of exposures (see log of
observations).

\section{Results}
\label{sect:results}

\subsection{Continuum visibilities}

The continuum visibilities computed with the different algorithms and
selection schemes are summarized in the right column of Table \ref{tab:V}.
We find discrepancies between the two methods that we are not able to
explain in this very early stage of the VLTI and AMBER. These
discrepancies might be related to the presence of vibrations in the
VLT UT coud\'e trains as mentioned in Sect.~\ref{sect:obs}, or to
remaining imperfection in the estimation or subtraction of the
biases.  We propose to take the following continuum visibility with a
relatively large error: $\Vcont=0.50^{+0.08}_{-0.10}$.
This confirms that the environment of \mwc\ is spatially resolved at
the level of a few milliarcseconds \citep{Eis2004}. The uniform disk
diameter corresponding to the visibility measured is
$7.0\pm0.9\,\mas$, corresponding to $1.75\pm0.23\,\AU$ at $250\,\pc$.

\subsection{Spectral variation of the visibilities}
\label{sect:visline}

\begin{figure}[t]
  \centering
  \includegraphics[width=0.9\hsize]{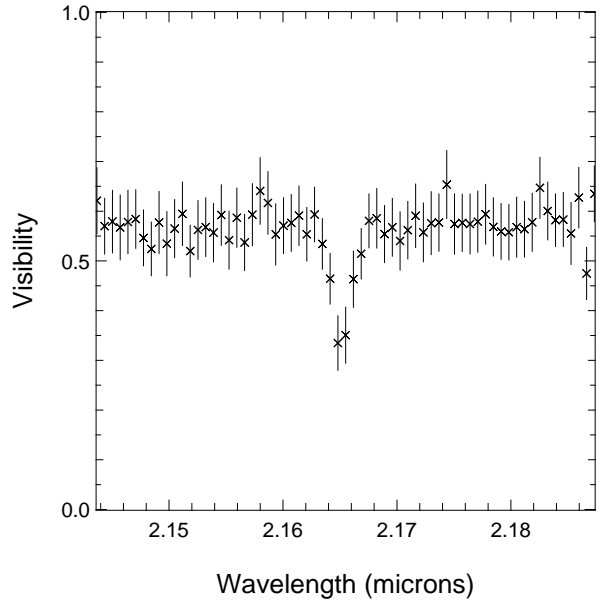}
  \caption{Spectral dependence of the visibility as measured with
    AMBER for \mwc\ around the \Brg\ line.}
\label{fig:specvar}
\end{figure}
All methods and selection schemes give consistent visibilities for the
\Brg\ emission line at \lBrg: $\Vline=0.33 \pm 0.06$ (see Table
\ref{tab:V}). Fig.~\ref{fig:specvar} shows the variation of the
visibility with wavelength for the $31\,\ms$ data. For this figure,
the continuum and line visibilities were calculated using the P2VM
and the best exposure selection criterion.

The visibility from the pure \Brg\ emission must be
corrected from the influence of the continuum by the following
relationship:
\begin{equation}
  \label{eq:vis}
  V_{\Brg} = \frac{\Fline\Vline - \Fcont\Vcont}{\Fline-\Fcont}
\end{equation}
where $F_{\Brg}=\Fline-\Fcont$. Since the ratio of the line
flux to the continuum flux is $\Fline/\Fcont=2.2$, the visibility of
the region emitting the \Brg\ line is $V_{\Brg}=0.19\pm0.03$,
corresponding to a uniform disk diameter of $9.9\pm0.3\,\mas$ or
$2.5\pm0.08\,\AU$ at $250\,\pc$.  Therefore, the size of the \Brg\
emitting region is 40\% larger (in terms of uniform disk
  diameter) than the size of the region contributing to the continuum.

\subsection{ISAAC spectrum}

\begin{figure}[t]
  \centering
  \includegraphics[width=\hsize]{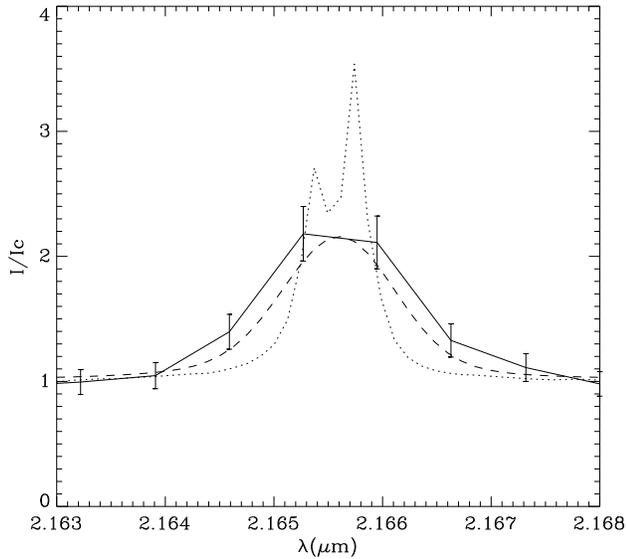}
  \caption{Comparison of \Brg\ observed with AMBER (solid line) and
    ISAAC (dotted line). The dashed line corresponds to the ISAAC
    spectrum convolved by the response function of} AMBER.
\label{fig:brgia}
\end{figure}
The AMBER spectrum at medium resolution ($\mathcal{R}\approx1500$)
contains the \Brg\ line.  Its profile has been compared to the ISAAC
high spectral resolution ($\mathcal{R}\approx8900$) spectra obtained
at nearly the same epoch in May and July 2004 respectively. A
comparison between the two lines is given in Fig.\,\ref{fig:brgia}.
The equivalent width of the AMBER spectrum is 2.14\,nm and is
compatible with the 1.58\,nm measured from the ISAAC spectrum given
the spectral variability of \mwc.  We note that the resolved infrared
\Brg\ emission lines presented in \citet{MD1994} are symmetric
rather than double-peaked.

The ISAAC spectral resolution is larger than the one achieved with
AMBER.  Nevertheless, in order to compare our two data sets the ISAAC
line profile was convolved with a Gaussian function corresponding to
the AMBER spectral resolution. The shape of both profiles are quite
similar (see Fig.\,\ref{fig:brgia}) but since the higher spectral
resolution of the ISAAC spectrum provides more details on the
kinematics within the circumstellar envelope, we have used, in the
following, the ISAAC \Brg\ line profile in order to constrain the
outflowing wind model (Sect.\,\ref{sect:wind}).

\section{Modeling}
\label{sect:models}

In this section we present the modeling of the large body of
interferometric, spectroscopic and photometric data that exists for
\mwc. The modeling is done by applying two different codes, one for an
optically thick disk and one for a stellar wind. The disk code is
designed to model the continuum radiation, whereas the stellar wind
code reproduces the strong emission lines.
\begin{figure}[t]
  \centering
  \includegraphics[width=\hsize]{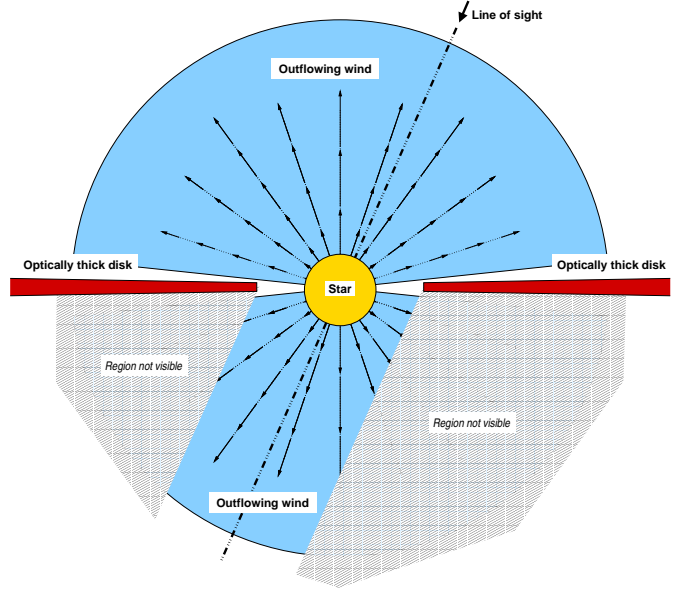}
  \caption{Sketch of the model including an optically thick disk and
    an outflowing wind (edge-on view). The receding part of the wind
    is only partly visible because of the screen made by the optically
    thick disk.}
  \label{fig:model}
\end{figure}
Figure \ref{fig:model} represents a sketch of the combined model, where the
optically thick disk and the outflowing wind are spatially independent.

\subsection{Continuum radiation: optically thick disk}
\label{sect:disk}

\begin{figure*}[t]
{
  \begin{center}
    \includegraphics[width=0.4\hsize]{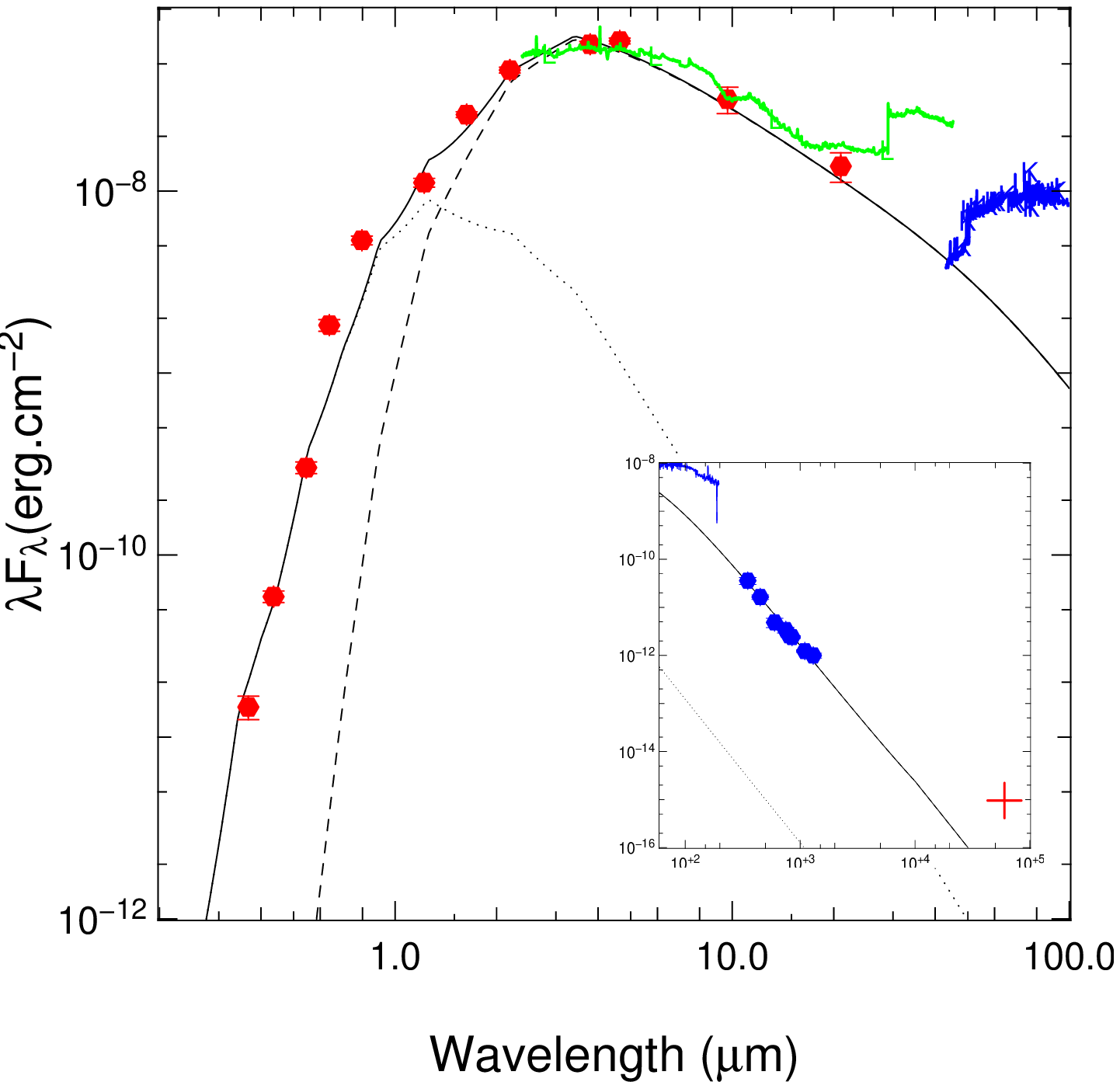}
    \includegraphics[width=0.38\hsize]{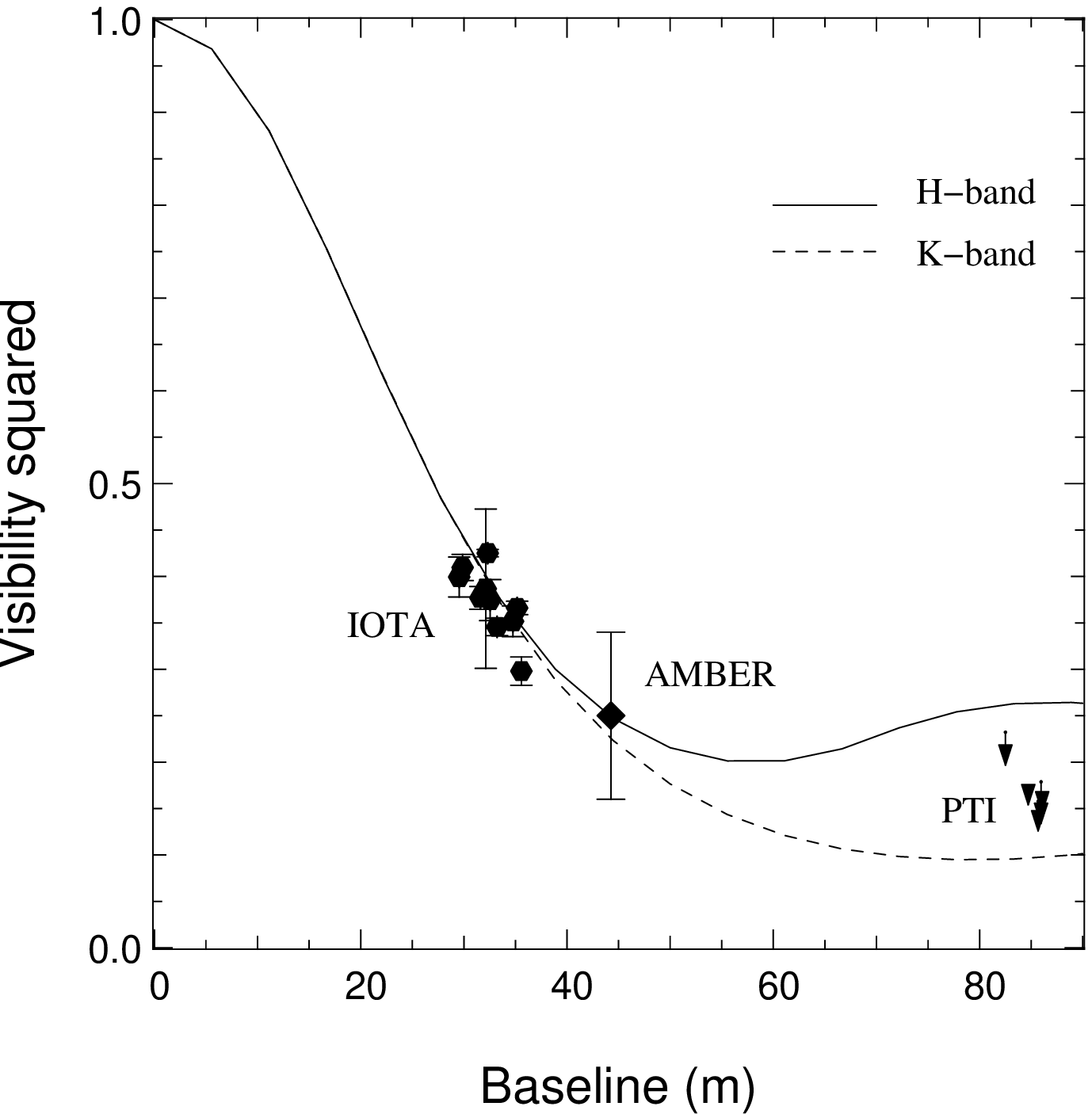}
    \caption{Result from the optically thick disk model. {\it Left
        panel:} observed and modeled SED for \mwc. The full dots are
      the continuum measurements taken from \citet{Pez1997}, also
      included are the ISO SWS/LWS spectra.  Dotted line is the star,
      dashed line the accretion disk, and the full line the resulting
      total flux of the model. {\it Right panel:} resulting best-fit
      model radial squared visibilities compared with continuum
      squared visibilities observed with AMBER, IOTA and PTI. Full
      line and IOTA data are in the $H$-band, dashed line and
      AMBER/PTI are in the $K$-band.  PTI values are upper limits.}
\label{fig:sed}
\end{center}
}
\end{figure*}

The disk model is based on the simple model already used by
\citet{MB1995} and \citet{Mal2005}. This disk model consists of an
axisymmetric radial analytic disk structure which is heated both by
stationary accretion and stellar irradiation. The disk is in
hydrostatic equilibrium and non self-gravitating. The accretion flux
is following the standard power law for a viscous disk
\citep{LBP1974,SS1973}. The emitted continuum flux is produced by the
emission of optically thick but geometrically thin black-body
radiating rings. It produces an SED, and its spatial distribution can
be Fourier transformed, which reduces to a Hankel transform for
  the radial symmetry of a disk, to obtain interferometric
visibilities.

The model effectively has five free parameters: the accretion rate
$\Macc$, the inclination $i$, the position angle $\theta$, and the
distances of the inner and outer edge of the disk, $\Rin$ and $\Rout$
respectively. Modeling is strongly constrained by the accurately
determined properties of the central star by \citet{Drew1997}:
a stellar radius of $6.12\,\Rsun$, a mass of $10\,\Msun$, an effective
temperature of $23,700\,\K$ for a distance of $250\,\pc$ with a visual
extinction of $\Av=8\,\Mag$. The outer edge of the
disk is strongly constrained by the mm/submm data point:
\citet{Man1994} finds an outer disk radius of $\simeq 60\,\AU$. In
short, simultaneous fitting of the SED and the visibilities lead to
balancing out the mass accretion rate, inner disk radius and the
inclination. In practice, this is done iteratively by first fitting the
SED, with $i$ and $\theta$ set to zero. We deem the $R$- and $I$-band
continuum measurements not reliable given the very strong \Ha\ 
emission; these two are therefore not taken into account in the fit.
In the first step we obtain \Macc\ and \Rin.  These are then used in a
separate fit of the visibilities as function of $i$ and $\theta$. The
results are used again for the first step and this procedure is
repeated until convergence. In Table\,\ref{tab:accpar} we list the
obtained best-fit model values.  The modeled SED and visibilities are
found to be in good agreement with the observed values, as shown in
Fig\,\ref{fig:sed}.

\begin{table}
 {
  \begin{center}
    \caption[]{The best-fit accretion disk model parameters found by
      simultaneous fitting of SED and visibilities. }
    \begin{tabular}{ll}
      \hline
      \hline
      Accretion rate ($\Macc$)  &$0-1\times10^{-5}\,\MsunPyr$\\
      Inner radius ($\Rin$)     &$0.5\pm0.1\,\AU$\\
      Outer radius ($\Rout$)    &$55\pm5\,\AU$\\
      Inclination ($i$)         &$15\pm5\degrees$\\
      Position angle ($\theta$) &$56\pm7\degrees$\\
      \hline
 \end{tabular}
\label{tab:accpar}
\end{center}
}
\end{table}

We probed the sensitivity of these fits by varying the central star
parameters, according to the uncertainties given by \citet{Drew1997}.
They derived half a spectral subtype uncertainty, and a distance error
of 50\,pc. Fitting the SED using the extreme values for the central
star, we find the uncertainties quoted in Table\,\ref{tab:accpar}.
Especially the mass accretion rate is far from well determined. If the
central star would be of type B2 at a distance of 200\,pc, the
required mass accretion rate is only $\sim 10^{-7}\,\MsunPyr$. 

\subsection{Emission lines: optically thin outflowing wind}
\label{sect:wind}

In our model, the emission lines are produced in a circumstellar gas
envelope. In order to model this line profile and the corresponding
visibilities, we have used the SIMECA code \citep{SA1994,Stee1995}.
This code computes classical observables, i.e.  spectroscopic and
photometric ones but also intensity maps in Balmer lines and in the
continuum in order to obtain theoretical visibility curves.  The main
assumptions are that (i) the envelope is axisymmetric with respect to
the rotational axis, (ii) no meridian circulation is allowed, (iii)
the physics of the polar regions is well represented by a CAK type
stellar wind model \citep{Cas1975}. The solutions for all stellar
latitudes are obtained by introducing a parametrized model (power of
sinus function) constrained by the spectrally resolved interferometric
data. Depending on the value of the chosen terminal velocity at the
equator, the equatorial region can be dominated either by Keplerian
rotation or by expansion.

Since the SIMECA code has originally been developed to model the
circumstellar environment of classical Be stars, we had to modify the
code in order to interface SIMECA with the optically thick disk model
described previously. We have implemented three changes (the
  equations describing the wind model are recalled in
  Appendix~\ref{sect:app}):
\begin{enumerate}
\item The wind is no longer computed from the equator to the pole, but
  the computation occurs in a bipolar cone defined by a minimal angle
  allowing the disk to be present (see sketch in
  Fig.~\ref{fig:model}). The disk model tells us that the opening
  angle is between $1.8\degrees$ at the inner radius and up to
  $3.2\degrees$ at the outer radius. We used a minimum angle of
  $4\degrees$.  Therefore the equatorial terminal velocity corresponds
  to the terminal velocity at this minimal angle from the
  equatorial plane at the interface between the accretion disk and the
  stellar wind.
\item The disk hides the receding part of the wind. In
  Fig.~\ref{fig:model}, the part of the wind which is not visible from
  the observer is not taken into account in the outgoing flux.
\item Although the disk emission contributes less than $1\%$ compared
  to the star flux in the visible (i.e.\ also in the \Ha\ and \Hb\ 
  lines) and can be neglected, at \lBrg\ the disk emission is 6.4
  times larger than the stellar flux.  This contribution is taken into
  account in addition to the excess due to free-free emission from the
  outflowing gas when computing continuum normalized intensities and
  also for the computation of the visibilities.
\end{enumerate}
To be consistent with the disk model (although it might not be
  fully physical), we also use the same temperature
and rotational velocity laws, respectively $T(r)\propto r^{-3/4}$ and
a Keplerian rotation.

\begin{figure}
  \centering
    \includegraphics[width=0.9\hsize]{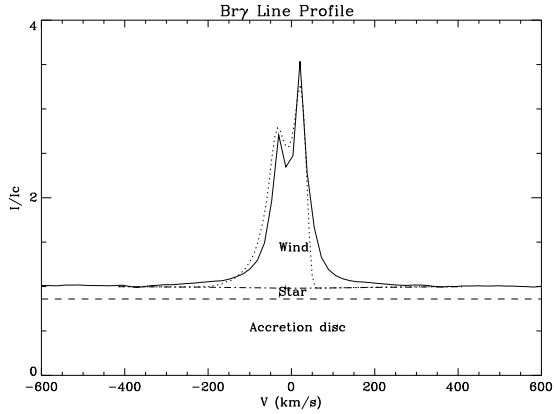}
    \caption{Double peaked \Brg\ profile observed by ISAAC (full line)
      and modeled with the outflowing wind model (dotted line). We
      have also plotted the cumulative contribution of the accretion
      disk (dashed line) and of the star (dashed-dotted).}
    \label{fig:lineprofiles}
\end{figure}
\begin{figure}[t]
  \centering
  \includegraphics[width=\hsize]{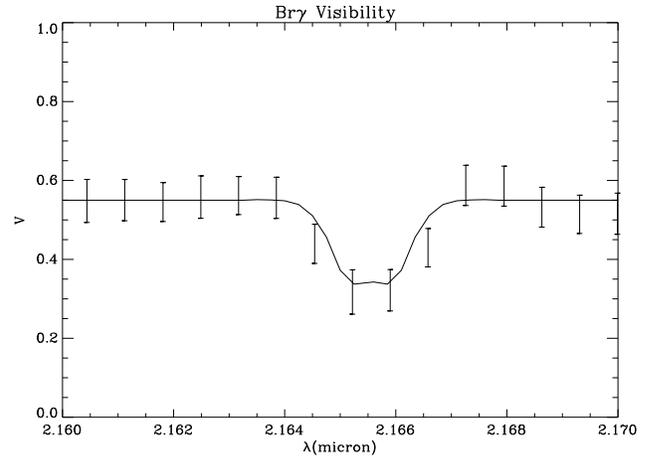}
  \caption{The visibility observed with AMBER (points with error bars)
    and the one obtained from the outflowing wind model (full line).}
  \label{fig:vis}
\end{figure}

\begin{table}
  \centering
  \caption[]{The best-fit model parameters for the outflowing wind
    model. Most parameters are self-explanatory (see
    Appendix~\ref{sect:app} for details). $C_1$ is the ratio
    between the polar and equatorial mass flux; $m_1$ is the
    exponent of the mass flux law as function of latitude; $m_2$ is
    the exponent of the latitude dependent terminal velocity law. }
  \label{tab:winpar}
  \smallskip
  \begin{tabular}{ll}
    \hline
    \hline
    Photospheric density               & $1\pm0.5\times10^{12}\,\text{cm}^{-3}$\\
    Equatorial rotational velocity    & $400\pm50\,\kms$\\
    Polar terminal velocity           & $600\pm50\,\kms$\\
    Terminal velocity above disk      & $70\pm20\,\kms$\\
    Polar mass flux                   & $3.2\pm0.2\times10^{-9}\,\MsunPyr$\\
    $C_1$                             & $0.25\pm0.05$\\
    $m_1$                             & $30\pm10$\\
    $m_2$                             & $10\pm2$\\
    Inclination angle ($i$)           & $25\pm5\degrees$\\
    \hline
  \end{tabular}
\end{table}
After running hundreds of simulations in order to constrain the
physical parameters of the wind and to test the geometrical and
kinematic hypothesis described later in Sect.~\ref{sect:geowind} , we
find a successful simultaneous fit to the ISAAC \Brg\ line profile
(see Fig.~\ref{fig:lineprofiles}) as well as \Ha, \Hb\ profiles
compatible with Drew's observations.  The best-fit model parameters
are given in Table\,\ref{tab:winpar}.  The outflowing wind model
reproduces successfully the AMBER measured drop in visibility across
the \Brg\ line as shown in Fig.~\ref{fig:vis}.

The computed \Ha\ and \Hb\ intensities are respectively 130 and 11.5
compared to 120 and 11 obtained by \citet{Drew1997}. This corresponds
to an intensity ratio $I_{\Ha}/I_{\Hb}$ of 11.3 compatible with the
Drew's ratio of 11 regarding the stellar variability and the
non-simultaneity of the measurements. Moreover we were able to
reproduce quite well the shape of these lines (i.e width) and the
double peaked emission of the \Brg\ line. The asymmetry of the two
peaks of the \Brg\ line is also successfully reproduced thanks to the
introduction in the SIMECA code of the opacity of the disk (point 2 of
SIMECA modifications). Nevertheless the agreement is not perfect in
the red wing of the profile probably due to our ad-hoc way of
interfacing of the wind and the disk.

The global shape of these lines and their intensities are very
sensitive to the inclination angle. We were able to determine the
inclination of $25\pm5\degrees$ from the fitting of the line profiles with
the SIMECA code. We note that the wind parameters given in
Table\,\ref{tab:winpar} are different from classical Be stars.
The mass flux ratio ($C_1$) is 0.25 whereas for Be stars the allowed
range is between 10 and 100.  Knowing that the density in the envelope
is proportional to the mass flux and inversely proportional to the
radial velocity we obtain that the gas density in the polar region is
twice larger than the one at the interface between the accretion disk
and the stellar wind. The highest mass loss occurs along the polar
  direction.

\section{Discussion}
\label{sect:discussion}

We have presented extensive existing data and new AMBER and ISAAC
data on the Herbig Be star \mwc. We have simultaneously modeled
interferometric and non-interferometric data. The continuum emission
is generated by an optically thick disk heated by stellar
irradiation and accretion, whereas optical and NIR emission lines
find their origin in an outflowing stellar wind.

The modeling presented in the previous section, although rather
successful, raises new questions on the physics of the circumstellar
environment of intermediate-mass young stars. In this section, we
discuss them by addressing first separate issues about the disk and
the wind, and then those raised by the combined model.

\subsection{Physical conditions in the disk}

Our continuum observations placed in the context of young stellar objects allow
us to bring new elements into the discussion of protoplanetary disk
models.

\subsubsection{Reprocessing versus accretion in \mwc}

The mass accretion rate is hardly constrained by the viscous accretion
disk model. In the first place because the viscosity in the accretion
disk is quite poorly understood, and therefore the quoted
accretion rate cannot be assessed with real certainty.  In the second
place, we obtain two orders of magnitude variation in the value of the
accretion rate, by varying the stellar parameters according to their
uncertainties (see Sect.~\ref{sect:disk}).

The ratio between the reprocessed energy from the stellar irradiation
and the viscous heating due to the accretion along the disk varies
from $4\times$ down to $3\times(\Macc/10^{-5}\,\MsunPyr)^{-1}$
from the inner part to the outer part of the disk. Therefore for
accretion rates lower than $10^{-9}\,\MsunPyr$, the reprocessing
energy is always dominant compared to the accretion energy. In fact
the transition between the two types of flux in the disk lies near a
$\Macc \simeq 3\times10^{-5}\,\MsunPyr$, for the most probable values
of the stellar parameters.

This means that the disk is predominantly a reprocessing disk. However
as pointed out by \citet[ see their Fig.~4]{LMM2003}, the emerging flux
might be dominated by the stellar heating and the vertical structure
by the viscous heating at least for the inner radii (less than a few AUs).

\subsubsection{Inner radius of the disk}

The best fit of our disk model is found with an inner disk boundary at
$0.5\,\AU$. This result has to be compared to the analysis of
\citet{MST2001}. With an assumption of $d=450\,\pc$, these authors
found a ring diameter of about $10\,\AU$ and an inner disk radius of
$0.9\,\AU$. Our result is therefore consistent with their disk model
since we have adopted the distance $d=250\,\pc$ of \citet{Drew1997}.

As already pointed out by other authors \citep{Eis2004,Mon2005},
the inferred inner disk radius for \mwc\ is not compatible with the
dust evaporation distance from the star. Using the disk model with the
"puffed up" inner rim \citep{DDN2001, IN2005}, the inner radius of the
disk would be located at about $3\,\AU$ from the star, even in the
most favourable hypothesis that the dust evaporation temperature is of
2000\,K. Moreover, \mwc\ seems not to be a special case, since this
behavior is common to the other two early HBe stars (Z~CMa A with a
spectral type B\,0 and V~1685 Cyg with spectral type B\,3) observed
with near infrared interferometers \citep{Mon2005}. One possible
explanation may be that the gas is the dominant source of absorption
in the inner part of the disk, preventing dust grains from evaporation
near to the star.


Using Rosseland gray opacities \citep{Ferguson2005}, we checked that
the disk is always optically thick both for an accreting disk
(accretion rate higher than $10^{-9}\,\MsunPyr$) or for a reprocessed
disk \citep[disk density at the disk edge higher than 1\% of
  the typical value of the solar nebula\footnote{Solar nebula density law used:
    $\Sigma_{SN} = 1.7 \times 10^3 \,\, (r/1\AU)^{-1.5}\,
    \text{g/cm}^2$};][]{Hayashi1981,Davis2005}.


Even with the last assumption, the structure of the inner region of
early HBe stars is not totally understood. The physical reason why the
optically thick gas disk should be truncated at $0.5\,\AU$ from the
star is not clear.

\subsubsection{Ionization of the disk} 
\label{sect:ionization} 

We estimate if the circumstellar disk is susceptible to
magnetohydrodynamical (MHD) instabilities, and thus if the disk is in
active accretion state by means of magneto-rotational instabilities
\citep[MRI;][]{BH1991}.  Given the large inner radius of $0.5\,\AU$,
it is not expected that the disk is accreting, otherwise the
region interior to the disk inner radius would quickly be filled.
Indeed, in the previously discussed hydrodynamic case, we saw that the
disk seems to be dominated by reprocessing.


Whether MHD instabilities in a given circumstellar disk operate or are
suppressed can be estimated with the magnetic Reynolds number
\citep[see e.g.][]{Gammie1996}. By setting the Reynolds number equal
to one, a threshold disk ionization fraction, $n_e/n_H$, of $3\times
10^{-13}$ is found. In this computation we adopted the same parameters
for the \mwc\ star-disk system as in the previous sections.  The
ionization fraction threshold is essentially determined by the
ionization rate and the particle density in the disk. Considering a
high ($10^{-5}\MsunPyr$) and a low ($10^{-9}\MsunPyr$) mass accretion
rate for \mwc, we can use Fig.~1 of \citet{IS2005} to deduce that the
required ionization rate in a dusty disk should range between
$10^{-14}$ and $10^{-15}\,\s^{-1}$ or between $10^{-10}$ and
$10^{-11}\,\s^{-1}$ respectively. Given that the cosmic ray ionization
rate is close to $10^{-17}\,\s^{-1}$, we conclude that cosmic rays
alone are insufficient to incite and/or sustain the required
fractional ionization for MRI to operate.

Since ultraviolet radiation is generally considered to be inefficient due
to the very small attenuation length, X-ray ionization may play an
important role in the ionization structure of the disk.  The very strong
X-ray activity reported for \mwc\ with ASCA \citep{Hamaguchi2000} should
be interpreted with care given the reasonably strong case for source
confusion \citep{Vink2005}. An inner disk radius of $0.5\,\AU$ is thus
probably not susceptible to MHD instabilities. We conclude thus that from
a MHD and a hydrodynamic point of view an inert reprocessing disk may
exist. We note however that if in the past the disk of \mwc\ was in fact a
magnetically active disk in which MRI operated, that then the concomitant
MRI turbulence could have sustained the ionization degree above the
threshold level even in the presence of dust \citep{IS2005}.

\subsection{Geometry of the wind}
\label{sect:geowind}

First, we investigate why the standard Be model of a rotating wind is
unsuccessful to reproduce the \mwc\ data. In this model
\citep{SA1994,Stee1995}, the \Brg\ emission would originate from a
rotating flattened envelope with no or little radial expansion. We can
test the consistency of this model by comparing the size of the
emitting region given by the kinematics revealed by the splitting of
the \Brg\ emission line into two peaks and the radius of this same
region inferred by the drop in visibility in this \Brg\ line.

On the one hand, the projected rotational velocity of the orbiting
material is given by the separation of the two \Brg\ peaks and is
proportional to the maximum extension of the emitting region
\citep{Huang1972,HK1984}. Using a rotational velocity law of the form
$v(r) =v_0\,(r/\Rstar)^{-x}$, where $v_0$ is the star rotational
velocity at the equator, $r$ is the distance to the star, and $x$ is
the exponent of the rotational velocity law, the separation between
the two peaks is proportional to $2v_0\sin i\,(r/\Rstar)^{-x}$.
Therefore the maximal radius of emission of the \Brg\ line is $\sim (2
v_0 \sin i/d)^{1/x}$, and the stellar parameters given by
\citet{Drew1997} and a Keplerian rotation yield $\sim140\Rstar$.
 
Since the typical radius of emission of the \Brg\ emission is
$\sim43\Rstar$ (see Sect.~\ref{sect:visline}), the size of $140\Rstar$
derived from the Keplerian rotation of a rotating stellar Be wind is
not consistent with our interferometric measurement.

In fact, in order to obtain an emitting zone radius compatible with our
interferometric measurement we need to assume a rotational
velocity of less than $250\,\kms$ or a rotation law far from the
Keplerian one. In total, we ran 50 models in our attempt to fit
the \Brg, \Ha, \Hb\ lines and the visibility in \Brg\ simultaneously using
this type of wind model. None of these models was able to fit correctly the
\Brg\ line profile displayed in Fig.\,\ref{fig:brgia} nor the intensity ratio
of the three hydrogen lines.  In addition, the \Ha\ and \Hb\ emission
lines require to form at least partially in a strong stellar wind to
reproduce the $600\,\kms$ line width in contrast with the \Brg\ line
displaying a $60\,\kms$ double-peaked profile. This is unlikely in the
same environment. 

\begin{figure*}[t]
  \centering
  \includegraphics[width=0.9\hsize]{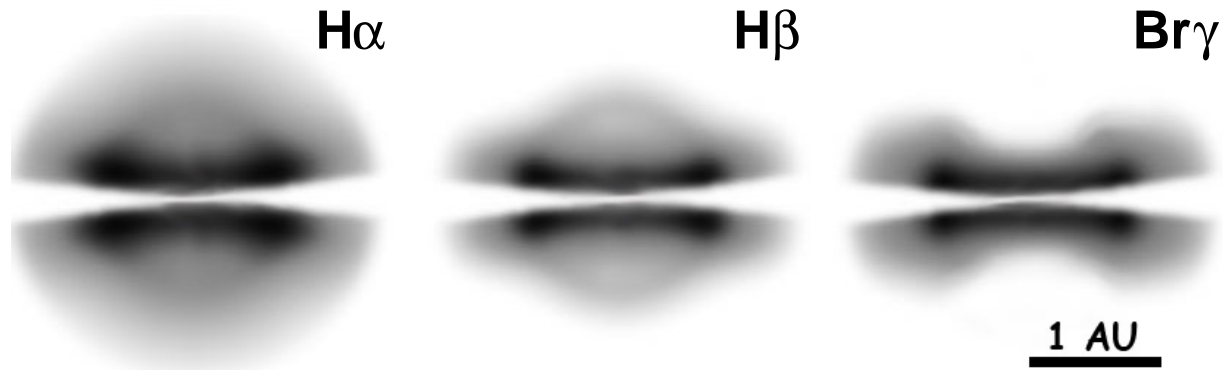} 
  \caption{Edge-on intensity maps of the wind in the computed \Ha,
    \Hb, \Brg\ lines.}
  \label{fig:windmaps}
\end{figure*}
The outflowing stellar wind coupled with an optically thick disk gives
a better explanation of the kinematical and geometrical morphology of
\mwc. In Fig.~\ref{fig:windmaps}, we present the intensity maps of
\mwc\ computed with our model seen edge-on in the three hydrogen lines
in order to better localize the region of emission of the lines. The
emission from $H\alpha$ and $H\beta$ originates from a large and
somewhat spherical region where the velocity can reach up to
$600\,\kms$.  The \Brg\ line is confined into a narrower region just
above the optically thick disk where the velocity is dominated by the
disk Keplerian rotation and a terminal velocity of $70\,\kms$ (see
parameters in Table~\ref{tab:winpar}). In fact \citet{MD1994} have already
pointed out that the \Bra/\Brg\ line flux ratio shows an increase at
low velocity that cannot be interpreted as an outwardly-accelerated
wind model.

\subsection{Disk and wind interaction in \mwc}

\begin{figure}[t]
  \centering
 \includegraphics[width=0.6\hsize]{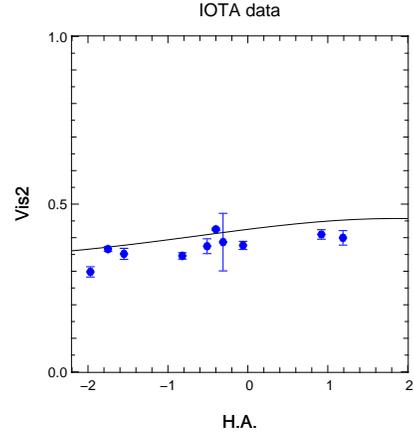}  
  \caption{Continuum $H$-band squared visibilities obtained with the disk
    model compared to the IOTA visibilities observed during the
    transit of the object over the sky with the IOTA S15N35 baseline.}
  \label{fig:iotavis}
\end{figure}
Even if the disk and the wind have been modeled separately, these two
physical phenomena need to be compared at least at the parameter
level. We find a good correspondence between the inclinations deduced
from the disk model ($i=15\pm5\degrees$ from the visibility
  fit) and from the wind model ($i=25\pm5\degrees$ from the
  asymmetry of the \Brg\ line peaks). The disk inclination is
strongly constrained by the IOTA visibilities spread out over a large
range of hour angle as shown in Fig.~\ref{fig:iotavis}.  The wind mass
loss rate is smaller than the maximum disk accretion rate by
several orders of magnitude, and therefore compatible with
most disk/wind theories but also with most observations of outflows
and accretion activities.  As shown in Sect.\,\ref{sect:ionization},
the inner radius is hardly ionized and thus will not contribute to the
hydrogen lines, confirming our choice to model \Brg\ only in the wind.

\begin{figure*}[t]
  \centering
  \includegraphics[width=0.3\hsize]{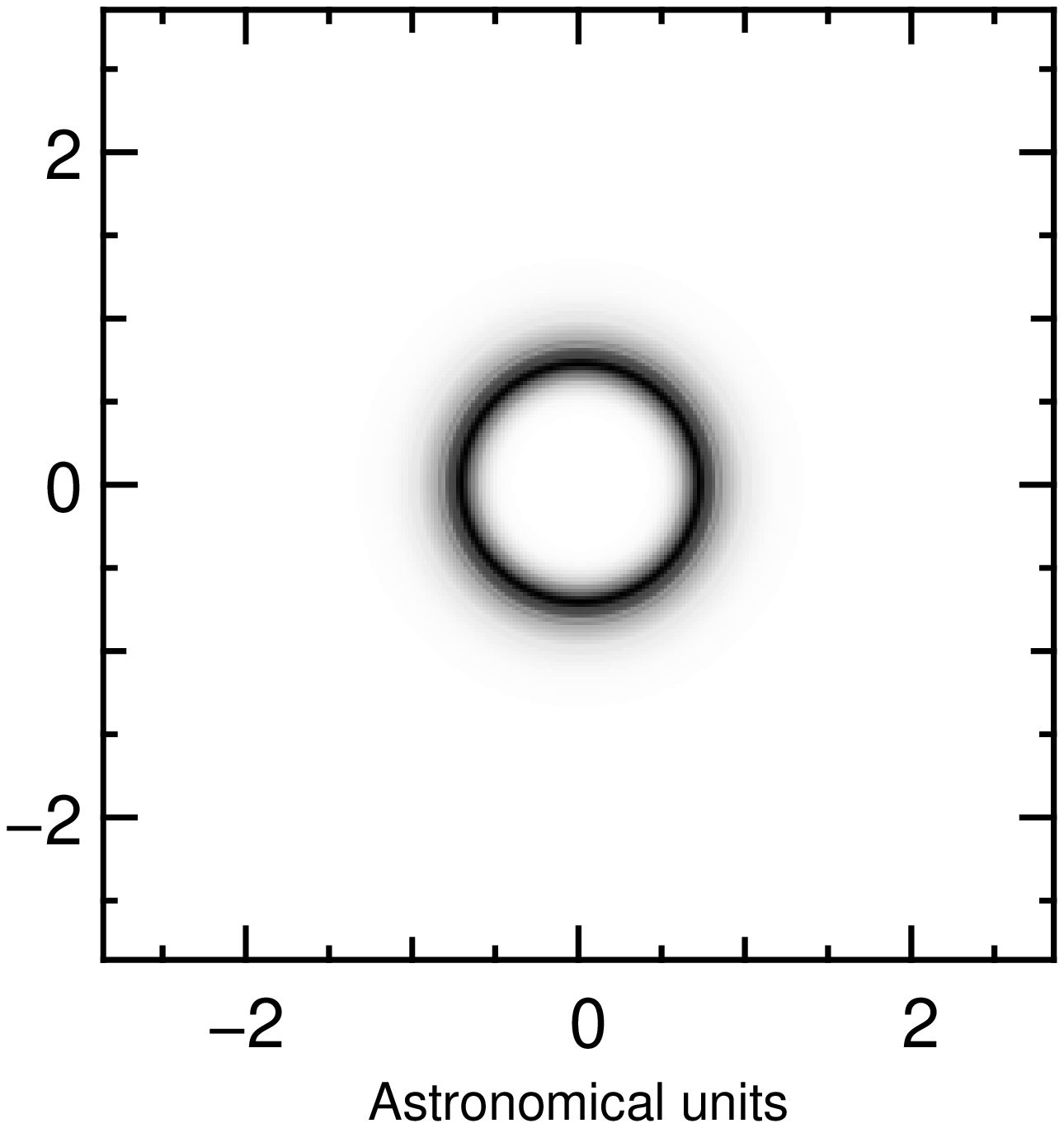} 
  \hfill
  \includegraphics[width=0.3\hsize]{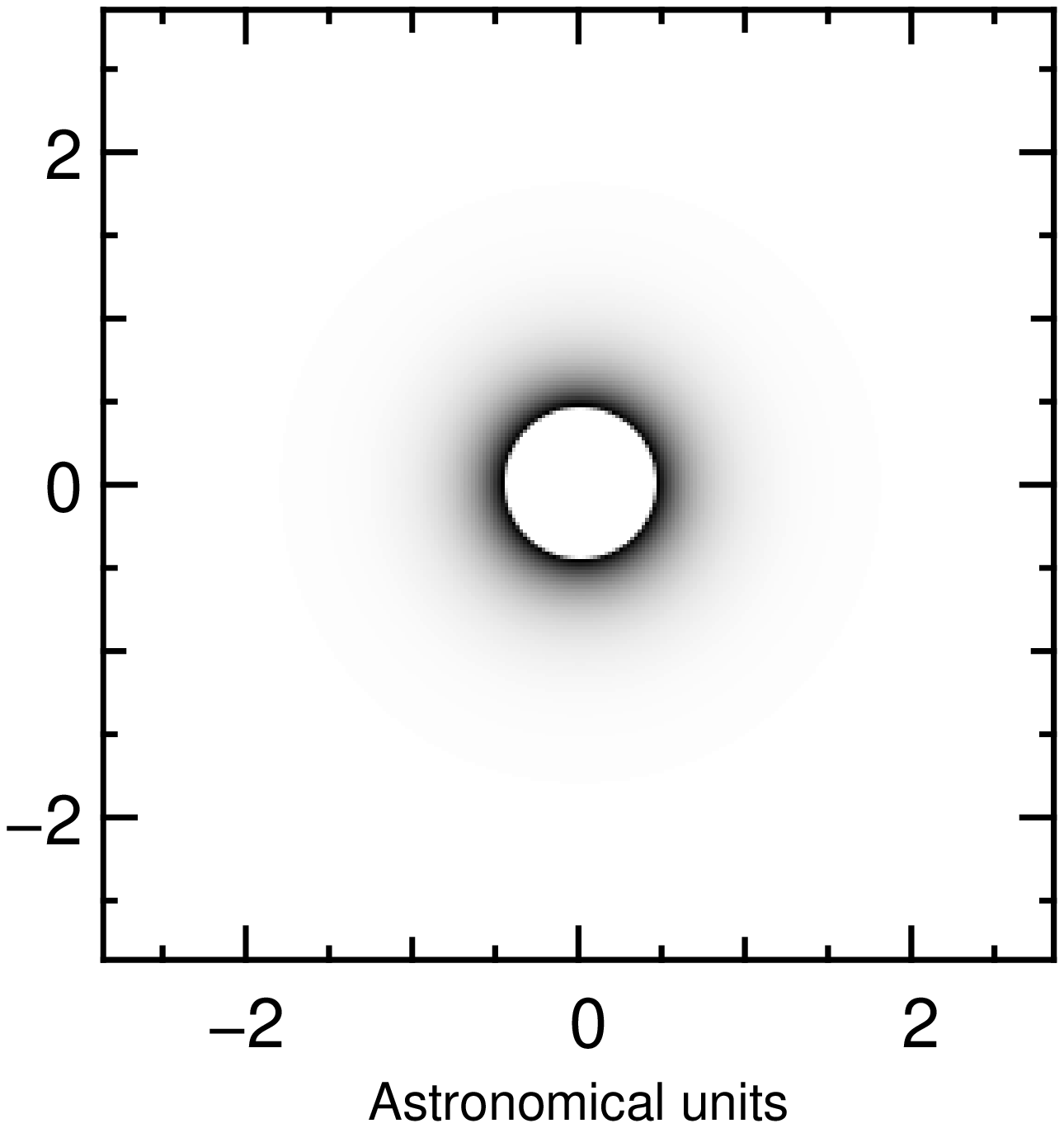}
  \hfill
  \includegraphics[width=0.3\hsize,height=0.32\hsize]{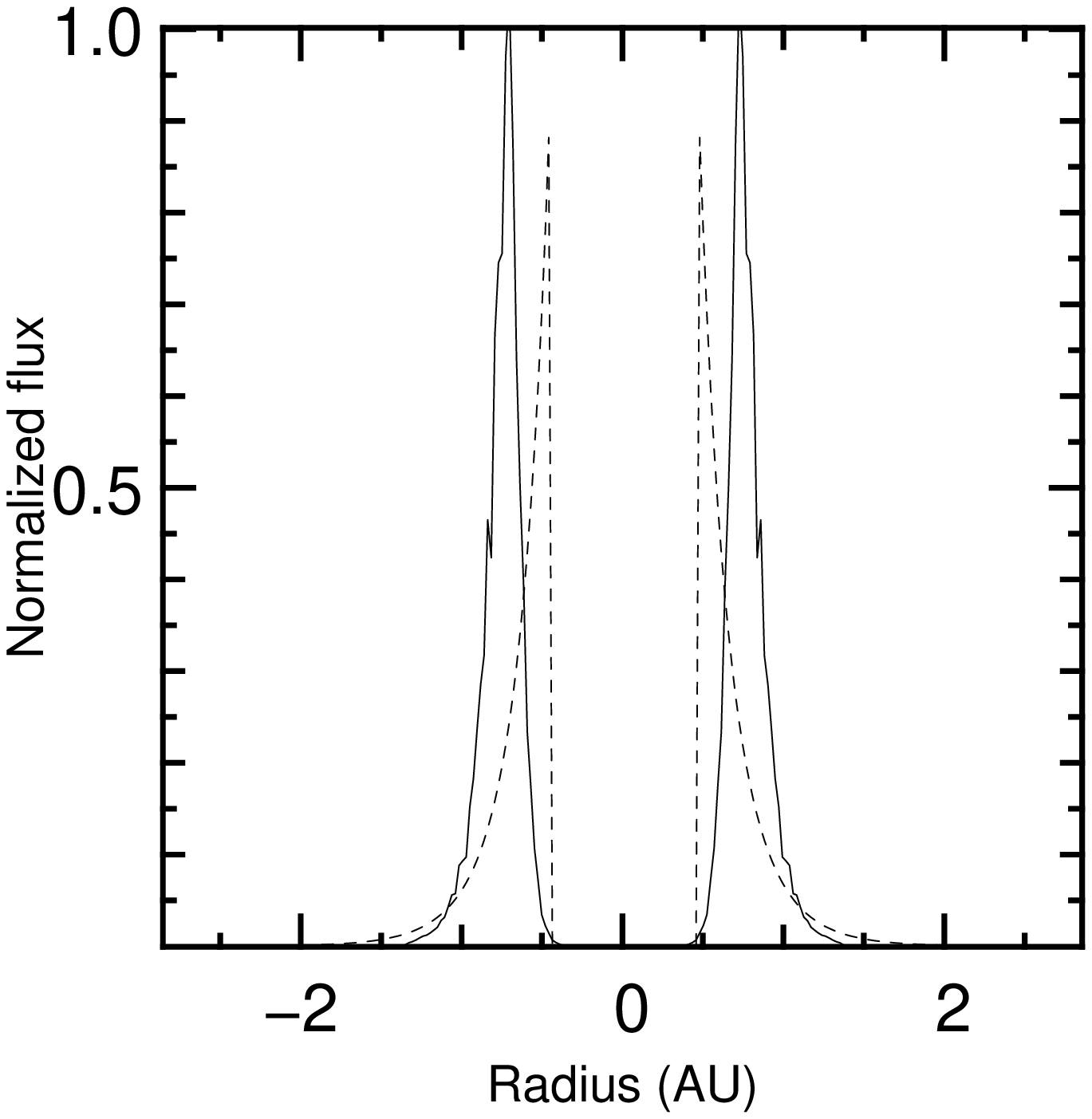} 
  \caption{Pole-on intensity maps of the wind \Brg\ emission (left
    panel) and of the $K$-band disk continuum emission (center panel).
    Right panel shows a radial cut of these intensity maps with the 
    \Brg\ wind in solid line and the continuum disk in dashed line.}
  \label{fig:intmaps}
\end{figure*}
Figure \,\ref{fig:intmaps} shows the pole-on intensity maps of the disk
model in the continuum and of the wind in the \Brg\ line, as well as
their respective intensity profile. This is a graphical explanation of
the visibilities observed by AMBER: the wind angular extension in the
\Brg\ line is larger than the disk apparent size, and therefore the
visibility is smaller within the line.

Can the result obtained with these observations constrain the nature
of the wind? We recall that in protoplanetary disks, two main classes
of disk wind models have been proposed depending on the geometry of the
magnetic field lines.
\begin{itemize}
\item The disk wind initially proposed by \citet{BP1982}
  and studied in detail for example in \citet{CF2000}. In this
  theory the magnetic field is attached to the disk and the particles
  which accrete to the star may be launched into a wind eventually
  collimated at a certain distance of the star.
\item The X-wind of \citet{Shu1994} has a different geometry and
  originates at the corotation radius of the disk where all magnetic
  field lines are localized in the equatorial plane.
\end{itemize}
With the present state of data, we cannot even distinguish between a
stellar wind and such disk winds since we are unable to recover a
precise kinematical map of the wind. We note however that our model
with Keplerian rotation does not respect the conservation of angular
momentum and can only be sustained if additional MHD energy is
injected in the wind. More spectral resolution with AMBER will help to
answer the question, especially using the $10,000$ spectral resolution
mode.

In conclusion, we can claim that the models of disk and wind are
compatible and the combination of them is probably close to
the reality. A complete and self-consistent modeling of the
environment is out of the scope of this paper but would allow to
better constrain the physical processes in action both in the
disk and the wind, like accretion and ejection, at least from
the observational point of view.

\subsection{Inclination discrepancy}

The modeling presented in this paper does a reasonably good job in
reproducing nearly all the observational data and produces fiducial
physical parameters for the circumstellar environment of \mwc. However,
we derive an inclination of $\sim 20\degrees$ for the system, which
is not consistent with a near edge-on orientation as proposed by
\citet{Drew1997}. The latter is inferred from the photospheric lines
that indicate a $350\,\kms$ projected rotational velocity. An
inclination of $20\degrees$ would lead to a rotation above the
break-up velocity.

The controversy becomes even more pronounced given the \Ha\ 
spectro-polarimetric results by \citet{OD1999}. These
authors find no effect across the line, a result consistent with a
symmetric emission zone for \Ha. On the contrary, the morphology of
the radio emission \citep{Drew1997} favors a bipolar interpretation
and again would point to a high inclination of the system, although
not completely inconsistent with a symmetric geometry on smaller
scales \citep{OD1999}.

A possible reconciliation of the seemingly contradictory observables
and derived physical parameters is a tidally induced warped geometry
for the circumstellar disk \citep{TerquemBertout1993} due to a stellar
companion located at a distance of a few hundred AU. Recently,
\citet{Vink2005} reported the presence of a close-by late-type
companion at a projected distance of 850\,AU. By converting the
$H$-band magnitude of the \citet{Vink2005} companion star to a
spectral type A2\,V and by calculating the steady-state deformation of
the \mwc\ circumstellar disk due to the star at 850\,AU, \citep[Eq.~25
of][]{TerquemBertout1993}, we find a deformation of less than $1/1000$
even at the edge of the disk at $60\,\AU$. Clearly this star would not
be the right candidate, but we feel that the possibility of a tidally
deformed disk may not yet be ruled out given the many companion stars
found near \mwc\ \citep{TPN1999,Vink2005}. An even closer companion
can not be excluded, a hint of which may be the CHANDRA X-rays found
coincident with the position of \mwc.
   
\section{Summary and perspectives} 
\label{sect:summ}

We have presented first spatially and spectrally resolved observations
of the disk/wind interaction in the young stellar system \mwc\ with
the VLT interferometer equipped with the instrument AMBER. The unique
spectral capability of AMBER has allowed us to measure for the first
time the visibility in the \Brg\ emission line in a young stellar
object and to compare it to the continuum visibility. The continuum
visibility in the $K$-band drops from 0.50 to 0.33 in the \Brg\
emission line of \mwc. The spectrum obtained with AMBER is consistent
with a double peaked spectrum observed with ISAAC on the VLT, where
the peaks are roughly separated by approximately $60\,\kms$.

We have successfully modeled the circumstellar environment of \mwc\ 
using an optically thick geometrically thin disk and an outflowing
stellar radial wind having an increasing outflowing velocity starting
from the surface of the disk up to the pole. This combined model is
able to reproduce many observational features like the continuum
visibilities measured by AMBER as well as the visibilities in the
\Brg\ line, together with the shape of the SED over more than three
orders of magnitude of the wavelengths, the broad-band visibilities
obtained by other infrared interferometers as
well as the \Ha, \Hb, and \Brg\ line profiles. 

We have discussed our result in the light of more sophisticated
models. We showed that the inner radius is not determined by the dust
sublimation distance, and is unlikely ionized by cosmic rays only.
The disk flux is mainly driven by the stellar reprocessing although we
cannot rule out that the accretion process may play a role on the
vertical structure.  We have shown that our AMBER observations
interpreted by our model predict emission of the \Ha\ and
\Hb\ lines in the polar regions whereas the \Brg\ emission arises
mainly from the region just above the surface of the disk. AMBER
continuum and line observations point both toward an system
inclination of approximatively 20\degrees.

We are not yet able to constrain the exact nature of the wind and the
type of connection with the disk, but we expect that future
observations with AMBER will bring new pieces in the understanding of
\mwc. As a matter of fact, the vibrations in the VLTI UT coud\'e train should
be diagnosed and fixed in a short term, the VLTI equipped with a
fringe tracker, allowing AMBER to be operated at its highest spectral
resolution ($\sim10,000$) which will give new kinematical
information on this interesting and intriguing region around
\mwc. Another direction of investigation would be to carry out AMBER
observations with three telescopes, in order to measure the closure
phase with different baseline configurations, and therefore measure
possible departure from centro symmetry of the material around \mwc.

\begin{acknowledgements}
  These observations would not have been possible without the support
  of many colleagues and funding agencies. This project has benefited
  of the funding from the French Centre National de la Recherche
  Scientifique (CNRS) through the Institut National des Sciences de
  l'Univers (INSU) and its Programmes Nationaux (ASHRA, PNPS). The
  authors from the French laboratories would like to thanks also the
  successive directors of the INSU/CNRS directors.  A.~Isella,
  L.~Testi and A.~Marconi acknowledge partial support from MIUR grants
  to the Arcetri Observatory: \emph{A LBT interferometric arm, and
    analysis of VLTI interferometric data} and \emph{From Stars to
    Planets: accretion, disk evolution and planet formation}.
  A.~Marconi acknowledges partial support from INAF grants to the
  Arcetri Observatory \emph{Stellar and Extragalactic Astrophysics
    with Optical Interferometry}. We would like to thank also the
  staff of the European Southern Observatory who provided their help
  in the design and the commissioning of the AMBER instrument. We are
  grateful to R.~Millan-Gabet and J.~Eisner who kindly provided their
  IOTA and PTI measurements.
  
  This work is based on observations made with the European Southern
  Observatory telescopes. The commissioning data can be retrieved from
  the ESO Science Archive Facility. This research has also made use of
  the ASPRO observation preparation tool from the \emph{Jean-Marie
    Mariotti Center} in France, the SIMBAD database at CDS, Strasbourg
  (France) and the Smithsonian/NASA Astrophysics Data System (ADS).
  The data reduction software \texttt{amdlib} is freely available on
  the AMBER site \texttt{http://amber.obs.ujf-grenoble.fr}. It has
  been linked with the free software
  Yorick\footnote{\texttt{ftp://ftp-icf.llnl.gov/pub/Yorick}} to
  provide the user friendly interface \texttt{ammyorick}.
\end{acknowledgements}

\appendix
\section{Wind model equations}
\label{sect:app}

Following \citet{SA1994} and \citet{Stee1995}, the mass flux is
parametrized as:
\begin{equation}
  \Phi(\theta)=\Phi_{\rm pole}+[(\Phi_{\rm eq}-\Phi_{\rm pole}) \sin^{m_1} (\theta)],
  \label{eq:phi}
\end{equation}
where $m_1$ is the first free parameter which describes the variation
of the mass flux from the pole to the equator. The ratio between the
equatorial and polar mass flux is $C_1= \Phi_{\rm eq}/\Phi_{\rm *pole}$.
The values of $C_1$ are typically between $10^1$ and $10^4$
\citep{LW1987}. Equation~(\ref{eq:phi}) can be rewritten as:
\begin{equation}
  \Phi(\theta)=\Phi_{\rm pole}[1+(C_1-1)\sin^{m_1} (\theta)].
  \label{eq:phi2}
\end{equation}
The expansion velocity field is given by:
\begin{equation}
v_r(r,\theta)=V_o(\theta)+[V_\infty(\theta)-V_o(\theta)](1-\frac{R} {r})^{\gamma},
\end{equation}
with $\gamma=0.86$ and 
\begin{equation}
V_o(\theta)=\frac{\Phi(\theta)}{\rho_{0}}=\frac{\Phi_{\rm pole}[1+(C_1-1) \sin^{m_1} (\theta)]}{\rho_{0}}.
\end{equation}

The second free parameter $m_2$ is introduced in the  expression of the
terminal velocity as a function of the stellar latitude:
\begin{equation}
V_\infty(\theta)=V_\infty({\rm pole})+[V_\infty({\rm eq})-V_\infty({\rm pole})]\sin^
{m_2} (\theta). 
\end{equation}






\noindent Finally the density distribution in the envelope is given  by the equation of
mass conservation:

\begin{equation}
\rho(r,\theta)=\frac{\Phi(\theta)}{{(\frac{r}{R})}^2 v_r(r,\theta)}.
\end{equation}


\bibliographystyle{aa}
\bibliography{paper-mwc297.bib}

\end{document}